\documentclass[aps,physrev,superscriptaddress,reprint]{revtex4-2}

\usepackage{orcidlink}
\usepackage{lipsum}  
\usepackage{graphicx}
\usepackage{amsmath}
\usepackage{float} 
\usepackage{ulem} 
\usepackage{amsfonts}
\begin{document}

\title{The phenomenological renormalization group in neuronal models near criticality}

\author{Kaio F. R. Nascimento~\orcidlink{0009-0009-1083-0044}}
\affiliation{Departamento de Física, Centro de Ciêcias Exatas e da Natureza Universidade Federal de Pernambuco, Recife, PE, 50670-901, Brazil}

\author{Daniel M. Castro~\orcidlink{0000-0002-2761-2133}}
\affiliation{Departamento de Física, Centro de Ciêcias Exatas e da Natureza Universidade Federal de Pernambuco, Recife, PE, 50670-901, Brazil}

\author{Gustavo~G.~Cambrainha~\orcidlink{0009-0000-8153-305X}}
\affiliation{Departamento de Física, Centro de Ciêcias Exatas e da Natureza Universidade Federal de Pernambuco, Recife, PE, 50670-901, Brazil}

\author{Mauro Copelli~\orcidlink{0000-0001-7441-2858
}}

\affiliation{Departamento de Física, Centro de Ciêcias Exatas e da Natureza Universidade Federal de Pernambuco, Recife, PE, 50670-901, Brazil}
\email{mauro.copelli@ufpe.br}

\date{\today}

\begin{abstract}
The phenomenological renormalization group (PRG) has been applied to the study of scale-invariant phenomena in neuronal data, providing evidence for critical phenomena in the brain. 
However, it remains unclear how reliably these observed signatures indicate genuine critical behavior, as it is not well established how close to criticality a system must be for them to emerge. 
Here, we rely on neuronal models with known critical points to investigate under which conditions the PRG procedure yields consistent results. 
We show that the PRG method detects scaling behavior in neuronal models only within a narrow vicinity of the critical point, reinforcing the interpretations drawn from PRG results in experimental data.
We also demonstrate that time-binning choices can substantially affect the results and introduce a data-driven adaptive binning procedure to circumvent this issue.

\end{abstract}

\maketitle

\section{Introduction}

The concept of criticality in biological systems has emerged as a fundamental approach to understanding the dynamics of living organisms, with significant implications for neuroscience \cite{Linkenkaer01, Beggs03, Chialvo10, Shew13, Beggs2007criticality, Munoz2018colloquium}.
The critical brain hypothesis \cite{Chialvo10, Shew13, PlenzNiebur14, TomenHerrmannErnst2019, obyrne_how_2022, hengen2024criticality} proposes that operating near a phase transition optimizes the brain's ability to process, store, and transmit information. 
The observation of scale-free neuronal avalanches in slices of the rat cortex \cite{Beggs03} ignited a surge of research on the critical brain hypothesis, leading to a wealth of both theoretical \cite{deArcangelis06, Kinouchi06a, Levina07, Levina09, Fraiman09, Bonachela09, Bonachela10, Larremore11a, Costa2015soc, Campos2017correlations}  and experimental \cite{Beggs04, Haldeman05, Stewart06, Plenz07, Pasquale08, Shew09, Petermann09, Ribeiro10a, Lombardi12, Tagliazucchi12, Yang2012maximal, PalvaPNAS13, shriki2013neuronal, Gautam2015, Shew15, Zhigalov2015, Fontenele2019criticality, lotfi2020signatures, lotfi2021statistical} evidence supporting the idea of criticality in the brain. Since then, research has expanded beyond avalanche dynamics, incorporating a broader range of statistical methods to explore the potential critical state of the brain \cite{Dahmen2019second, sompolinsky2022spectrum, munoz2024}.

Among these, the phenomenological renormalization group (PRG) \cite{bradde_pca_2017, meshulam_coarse_2019,bialekreview2025} has emerged as a novel technique to study criticality signatures of complex systems. 
This coarse-graining framework, inspired by the renormalization group (RG), a theoretical cornerstone of critical phenomena, provides a model-independent approach to investigating scale-invariant behavior. 
While RG requires a well-defined model to study phase transitions and criticality, PRG analysis is model-free and works directly with empirical data, enabling the analysis of scale-invariant phenomena in high-dimensional datasets, such as neural activity. 
The PRG method is based on two  coarse-graining approaches: the first, in real space, consists in iteratively pairing the most correlated neurons~\cite{meshulam_coarse_2019}; 
the second is a momentum space transformation, in which data is  projected into a progressively coarser representation through the ranked eigenvalue spectrum of the covariance matrix~\cite{bradde_pca_2017}. 
These approaches allow one to identify scale-invariant observables as data is rescaled. 
One also expects a convergence to non-Gaussian fixed points in the case of critical systems and a tendency towards a Gaussian distribution for noncritical ones~\cite{bradde_pca_2017, meshulam_coarse_2019}.

PRG has been applied to a plethora of datasets, ranging from neuronal calcium fluorescence~\cite{meshulam_coarse_2019, munn_multiscale_org} and spiking activity~\cite{Morales2023scaling, castro_and_2024, cambrainha2025criticality} up to whole-brain fMRI recordings~\cite{ponce2023critical, castro2025interdependent}. 
In all cases, data has been time-binned. 
These studies have reported striking interpretations of scale invariance in neural data such as universal scaling across regions \cite{Morales2023scaling} and species \cite{munn_multiscale_org}, state-dependent scaling \cite{castro_and_2024}, and links between criticality and performance \cite{cambrainha2025criticality}. 
All of these conclusions rely on PRG as a diagnostic of non-trivial scale-invariant dynamics. Yet, despite such strong claims, it remains unclear whether such scaling signatures appear only in the vicinity of a critical point.

The first incursions into applying PRG to canonical models, such as the contact process and the Ising model, have been made by Nicoletti et al.~\cite{nicoletti_scaling_2020} and Ponce-Alvarez et al.~\cite{ponce2023critical}, respectively. However, a systematic investigation of how scaling emerges in momentum space as a control parameter is finely tuned across different dynamical phases is still lacking. Moreover, these studies focused on models that do not capture neuronal population dynamics and lacked key biological features, such as inhibition,  limiting their comparability with real neuronal data.

\begin{figure*}[t]
    \centering
    \includegraphics[width=\linewidth]{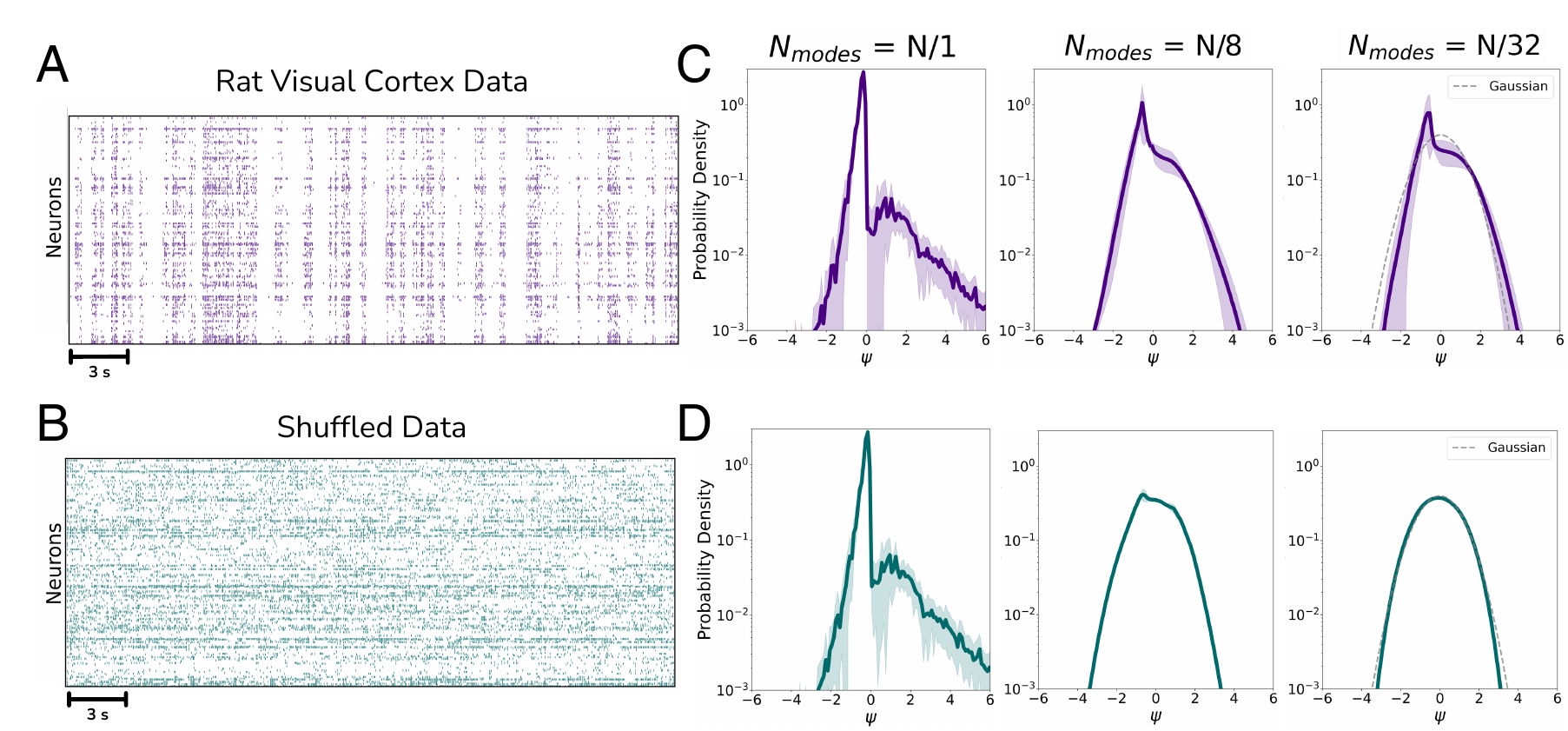}
    \caption{   (A) Raster plot of spiking activity recorded from urethane‐anesthetized rat visual cortex.  
    (B) Raster of shuffled data obtained by shuffling each neuron’s mean inter‐spike interval (same overall spike count and binning as A). 
    (C) Probability density of the normalized coarse-grained activity $\psi$ at successive PRG scales ($N_{\mathrm{modes}} = N/1, N/8, N/32$) for the real data, illustrating a deviation from Gaussian behavior. 
    The shaded regions represent standard deviation across 373 windows of 30 seconds~\cite{castro_and_2024}. 
    The dashed gray line represents a unit-variance Gaussian.
    (D) Same analysis as in (C) applied to surrogate data, illustrating convergence toward Gaussian statistics.}
    \label{fig:fig1}
\end{figure*}

Here we have a closer look at how the PRG scaling signatures are affected while a model is tuned away from the critical point.
As one sweeps parameter space from subcritical to supercritical phases in a spiking network model, the distribution of interspike intervals of can vary substantially, thus changing the relevant timescale under which network activity should be analyzed. 
We will show that this leads to problems when we use a fixed bin size across different dynamical phases. 
This kind of pre-processing issue has been addressed in avalanches~\cite{Capek2023parabolic, srinivasan2024recovery} and we do it here for PRG analysis.

We simulate two models with exactly known critical points, both belonging to the mean-field directed percolation (MF-DP) universality class (which is widely employed to model experimentally observed neuronal data~\cite{Beggs03}). 
These models display a phase transition from an absorbing state to an active state. 
The first model is a spiking cellular automaton, where the control parameter is the branching ratio, defined as the average number of cells activated by a single active cell in the following time step. 
The second model is a stochastic integrate-and-fire network with inhibition, where the intensity of inhibition serves as the control parameter, modulating the excitatory-inhibitory interactions. 
We examine PRG signatures as the control parameters evolve from the absorbing to the active phase, analyzing the transition across the critical point.

Our approach consists of generating data by simulating the two models while gradually sweeping across parameter space, collecting data from both the subcritical and supercritical regimes. We then apply PRG analysis to examine how scaling signatures emerge near criticality. As we will see, correctly addressing the time-binning step of the data preprocessing is relevant to obtain consistent results.

\section{Methods}

In this section, we outline the PRG coarse-graining procedure, including both its real-space and momentum-space implementations. 
As shown in \cite{meshulam_coarse_2019} and \cite{bradde_pca_2017}, the former links Kadanoff spin-block coarse-graining to correlation-based network decimation, while the latter connects a field-theoretic RG framework to principal component analysis (PCA).
 We then describe the models analyzed, specifying the simulation parameters, time window selection, and how the time bin size affects our results.

\subsection{Phenomenological renormalization group}
\subsubsection{Momentum-space coarse graining}\label{sec:momentumrenorm}
A key aspect of the PRG framework is the momentum space transformation. This transformation relies on the spectrum of the covariance matrix to define a coarse graining procedure in terms of its dominant eigenmodes. We begin with the covariance matrix:
\begin{equation}
    C_{ij} = \langle \varphi_i \varphi_j \rangle - \langle \varphi_i \rangle \langle \varphi_j \rangle,
\label{eq:covariance}
\end{equation}
where 
$\varphi_i (t)$ is the sequence of spikes of neuron $i$, i.e. a binary vector where 0 represents silence and 1 represents a spike. 
Here the angle brackets represent the temporal average. The eigenvalues $\lambda_1 > \lambda_2 > \dots > \lambda_N$ and corresponding eigenvectors $u_{\mu i}$ satisfy
\begin{equation}
    \sum_{j=1}^{N} C_{ij} u_{\mu j} = \lambda_{\mu} u_{\mu i}.
\end{equation}
Ranking these eigenvalues in ascending order up to some  $N_{\text{cutoff}}$, we can define the projector:
\begin{equation}
    \hat{P}_{ij}(N_{\text{cutoff}}) = \sum_{\mu=1}^{N_{\text{cutoff}}} u_{\mu i} u_{\mu j}.
\end{equation}
This allows us to obtain the following set of transformed variables for a given $N_{\text{cutoff}}$
\begin{equation}
\label{eq:psi}
    \psi_i(N_{\text{cutoff}}) = Z_i(N_{\text{cutoff}}) \sum_{j=1}^{N} \hat{P}_{ij} (N_{\text{cutoff}}) (\varphi_j - \langle \varphi_j \rangle) ,
\end{equation}
where $Z_i(N_{\text{cutoff}})$ is a normalization that guarantees $\text{var}(\psi) = 1$. 
$\psi_i$ is the zero-mean, normalized coarse-grained version of $\varphi_i$, as we geometrically reduce the dimensionality of the projecting eigenspace: $N_{
\text{cutoff}} = N, N/2, N/4, \dots$~\cite{bradde_pca_2017}. Following this prescription, we may calculate the distribution of transformed variables at increasing coarse-graining levels, namely:

\begin{equation}
P(\psi) = \frac{1}{N}\sum_{i=1}^N \mathbb{P}[\psi_i(N_{\text{cutoff}})=\psi].
\label{MSdist}
\end{equation}

This coarse-graining process is illustrated in Fig.~\ref{fig:fig1}, where the method was applied to cortical spiking data from the primary visual cortex of rats anesthetized with urethane~\cite{castro_and_2024}. 
One observes an evolution to a non-trivial distribution $P(\psi)$ (Fig.~\ref{fig:fig1}C), whereas surrogate data (obtained by shuffling interspike intervals of each neuron, see Sec.~\ref{sec:surrogate}) leads to an approximately Gaussian distribution (Fig.~\ref{fig:fig1}D). Notice that, even if the activity distribution of the system is initially non-Gaussian, the coarse-graining procedure drives the distribution to a Gaussian form in the absence of scale invariance.

There are many ways to assess the gaussianity (a proxy for triviality) of $P(\psi)$. 
One possibility that has been employed previously is to
calculate 
its kurtosis, ${\kappa = \langle \psi^4 \rangle / \langle \psi^2 \rangle^2}$, and compare it with the kurtosis of the distribution of surrogate data. 
We refer to the difference as $\Delta \kappa$.
 The kurtosis of the distribution of coarse-grained activity $P(\psi)$ has been used as a metric for criticality in several recent works, involving applications of PRG in experimental data \cite{munn_multiscale_org,castro_and_2024, cambrainha2025criticality}.
Here we show that other conventional measures can also be equally informative. 
We calculate the Kullback-Leibler divergence (KL) of $P(\psi)$ with respect to a zero-mean unit-variance gaussian ${\mathcal{N}}$
\begin{equation}
    \mathrm{KL}(P(\psi)||\mathcal{N}) = \int P(\psi)\log_2 \left(\frac{P(\psi)}{\mathcal{N}}\right) d\psi\; ,
\end{equation}
which equals zero when $P(\psi)=\mathcal{N}$. 
We also calculate KL for $P(\psi_{surr})$ obtained for surrogate data (see section~\ref{sec:surrogate}). 
The difference between the two is $\Delta \mathrm{KL}$.
Nonetheless, kurtosis as well as other metrics can also be used, with similar results (see Appendix~\ref{alt_gauss}).

\subsubsection{Real-space coarse graining}
\label{sec:realspace}

Another proposed measure of critical behavior is given by the PRG exponents, obtained through real-space coarse-graining. This process involves successively combining the most correlated pair of neurons until no neuron is left unpaired. Initially, we have $N$ neurons, each represented by a binary variable $\sigma_i^{(1)} \in \{0,1\}$. 

At each coarse-graining step $n+1$, we merge the most correlated pair of variables into a new coarse-grained variable (or cluster), by way of a simple sum:  
\begin{equation}
    \sigma_i^{(n+1)} = \sigma_i^{(n)} + \sigma_j^{(n)}\text{.}
\end{equation}
Here the  index $j$ denotes the neuron maximally correlated with neuron $i$. We then repeat this process by selecting the next most correlated pair and merging them in the same way. After one coarse-graining step, we are left with $N/2$ clusters.  
After $k$ coarse-graining steps, the system will consist of $N/2^k$ clusters of size $C_{\text{size}}= 2^k$. 
As the system nears criticality, we expect to see some quantities scale with non-trivial power laws
as a function of $C_{\text{size}}$, such as the intracluster mean variance, the probability of silence, and the average autocorrelation time~\cite{meshulam_coarse_2019}.
As we shall see, the results from the momentum-space approach turn out to be more precise for the purposes of detecting scale invariance. 
Therefore, as an illustrative example, we restrict our real-space analysis to the intracluster mean variance, which follows a power law: 
\begin{equation} \label{alpha}
M_2 \propto C_{size}^\alpha\text{.}
\end{equation}
For uncorrelated units, we expect $\alpha = 1$ (central limit theorem), while for perfectly correlated units we find ${\alpha = 2}$. 
For a non-trivial (critical) case, we expect ${1 < \alpha < 2}$.

\subsection{Excitable cellular automata model}

To examine the extent to which PRG reliably identifies critical behavior, we apply it to an excitable cellular automaton model, where each element may spike due to the excitation of its neighbors~\cite{Kinouchi06a, Carvalho2021subsampled}. 
The model is defined on a random network topology, with each unit unidirectionally connected to $K$ neighbors. 

Each unit \( i = 1, \dots, N \) is characterized by \( n \) discrete states: \( s_i = 0 \) represents the resting state, \( s_i = 1 \) denotes the excited (spiking) state, and \( s_i = 2, \dots, n-1 \) correspond to refractory states. The probability that element \( i \) transitions from \( s_i = 0 \) to \( s_i = 1 \) as a result of a neighbor \( j \) being in the excited state at the previous time step is given by \( p \). 
Once an element is excited (\( s_i = 1 \)), its state progresses deterministically: $s_i(t+1)=(s_i(t)+1) \mod n$. 
The order parameter is the stationary density of active sites $\rho$, whereas the control parameter is the branching ratio $\sigma = Kp$. An MF-DP phase transition occurs at  $\sigma_c = 1$~\cite{Kinouchi06a, Carvalho2021subsampled}, above which $\rho$ departs from zero (Fig.~\ref{fig:fig3}B). 

\subsection{Spiking model with excitation and inhibition}
The second model to which we apply the PRG analysis is based on the network described by Girardi-Schappo et al.~\cite{girardi2020synaptic}. This model consists of excitatory and inhibitory neurons, each modeled as stochastic leaky integrate-and-fire units, connected in a complete network topology where every neuron is connected to all other neurons.

The membrane potential \(V^{E/I}_i(t)\) for  each excitatory \((E)\) and inhibitory \((I)\) neuron \(i\) evolves according to the following equation:
\begin{align}
V_i^{E/I}(t+1) = & \Bigg[ \mu V_i^{E/I}(t) + I_e + \frac{J}{N} \sum_{j=1}^{N_E} X^E_j(t)  \notag \\
& - \frac{gJ}{N} \sum_{j=1}^{N_I} X^I_j(t) \Bigg] \Bigg(1 - X^{E/I}_i(t)\Bigg),
\end{align}
where \(V^{E/I}_i(t)\) is reset to zero after each spike. 
The leak time constant is \(\mu\), \(I_e\) is the external current, and \(J\) represents the synaptic coupling strength. 
The binary variables \(X^E_j(t)\) and \(X^I_j(t)\) indicate whether neurons in the excitatory or inhibitory populations are firing \(X(t) = 1\) or not \(X(t) = 0\) at time \(t\). The parameter \(g\) is the inhibition-to-excitation coupling strength ratio.

We consider a network  with 80\% excitatory neurons 
 and 20\% inhibitory neurons, to emulate cortical data~\cite{Somogyi1998}
. The firing probability of each neuron is given by:
\begin{align}
\Phi(V) \equiv P(X = 1 | V) = & \; \Gamma (V - \theta) \Theta(V - \theta) \Theta(V_S - V) \notag \\
& + \Theta(V - V_S).
\end{align}
where \(\Gamma\) is the firing gain constant, \(\theta = 1\) is the firing threshold, \(\Theta(x>1) =0\) (null otherwise) is the step-function and \(V_S = \frac{1}{\Gamma + \theta}\) is the saturation potential.

For simplicity, we set \(\mu = 0\) since this choice does not affect the critical point of the model~\cite{girardi2020synaptic}. 
Like in the cellular automaton model, here the order parameter is the density $\rho$ of active sites. 
Choosing $g$ as our control parameter and fixing \(\Gamma = 0.2\) and \(J = 10\), the model exhibits an MF-DP phase transition at \(g_c =  1.5\). 
When \(g < g_c\), the system is in the excitation-dominated supercritical phase, and when \(g > g_c\), it enters the inhibition-dominated subcritical absorbing state. 
At the critical point, excitation and inhibition fluctuations are dynamically balanced.

\subsection{Phenomenological renormalization group applied to models}
\subsubsection{Simulation parameters}

We simulate both models using networks of \(10^4\) neurons, dividing the resulting activity into trials of \(5 \times 10^3\) time steps. 
For the sake of comparison with experimental data, we consider each time step to represent 1~ms (in both models). 
Each simulation begins with all neurons in a quiescent state, except for a single neuron that initiates network activity. To prevent the system from remaining in the absorbing state, we excite a single, randomly selected neuron (excitatory in the second model) whenever activity ceases.  
After completing the simulations, we randomly select $N_\text{prg} = $ 256 neurons to perform the PRG analysis. This subsampling emulates the limited observability inherent in experimental recordings, where only a small fraction of the neural population is typically accessible~\cite{Carvalho2021subsampled}.

\subsubsection{Binning the data}

We bin the data by taking the time series and dividing it into intervals of width \(\Delta t \). 
For each time bin, ${\varphi_i=1}$ if neuron $i$ spiked at least once within the bin ($\varphi_i=0$ otherwise, see Fig.~\ref{fig:fig3}A). 
The average active bin density is $\rho_{bin} = (T N)^{-1}\sum_i^N \sum_t^T \varphi_i(t)$, and naturally increases with $\Delta t$ (Fig.~\ref{fig:fig3}C). 
This will be important, since the PRG method is completely based on the correlations between $\{\varphi_i\}$. 
For $\Delta t$ too small (too large), most $\varphi_i$'s will be zeros (ones). 

\begin{figure}[htpb]
    \centering
    \includegraphics[width=\linewidth]{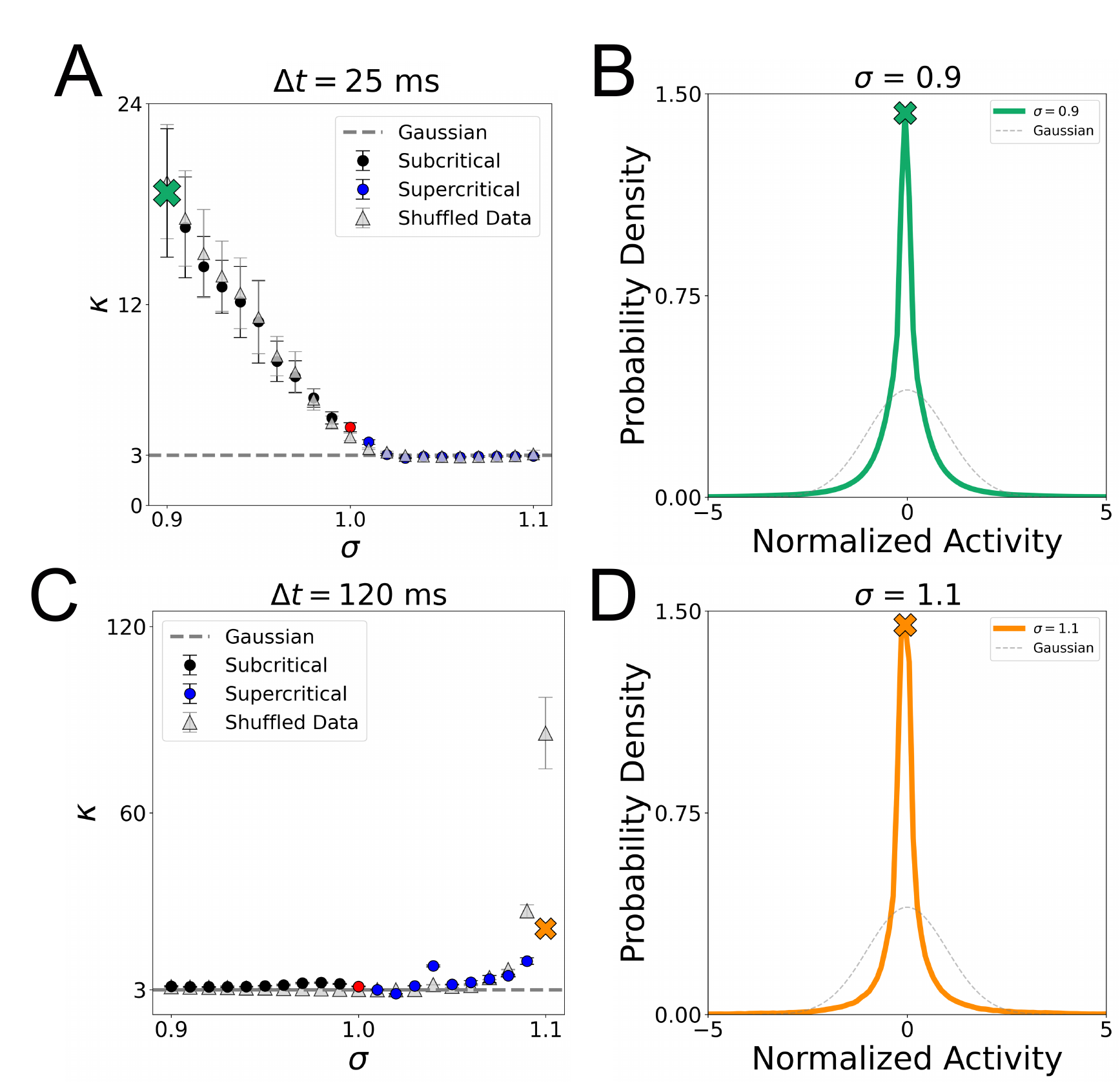}
    \caption{PRG analysis of the excitable cellular automaton model, illustrating the effect of time bin size (\(\Delta t\)) on kurtosis measurements. (A) and (B): \(\Delta t = 25\)~ms. (A) Kurtosis \(\kappa\)  attains high values even for subcritical coupling \(\sigma\), incorrectly suggesting critical behavior. (B) Probability density of normalized activity at \(\sigma = 0.9\) displays a non-Gaussian distribution due to low activity levels rather than genuine criticality. (C) and (D): \(\Delta t = 120\) ms. (C) A large bin size causes elevated kurtosis for supercritical \(\sigma\). (D) Probability density at \(\sigma = 1.1\) with high \(\kappa\) results from activity saturation. Dashed lines represent Gaussian distributions in (B) and (D);  triangles denote shuffled data.
}\label{fig:fig2}
\end{figure}

\begin{figure*}[htbp]
    \centering
    \includegraphics[width=\linewidth]{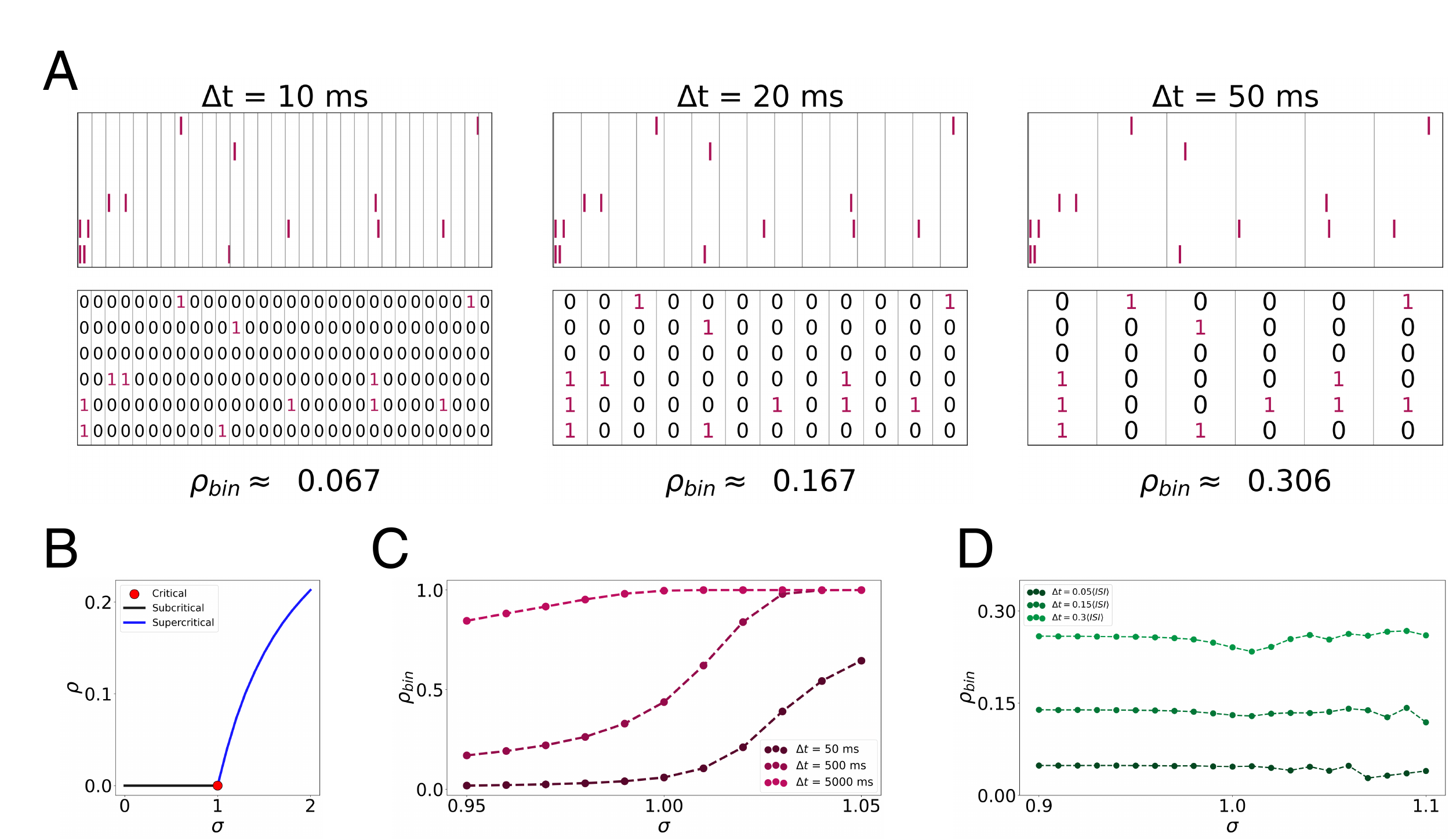}
    \caption{(A) Illustration of the effect of choosing a time bin on a  raster plot. Spiking activity is discretized using different bin sizes (\(\Delta t = 10\)~ms, \(\Delta t = 20\)~ms, and \(\Delta t = 50\)~ms), leading to different binary representations of neural activity. (B) Phase transition in the excitable cellular automaton model, showing the relation between the control parameter \(\sigma\) and the activity density \(\rho\). The system exhibits an absorbing phase  for \(\sigma < 1\) and an active phase for \(\sigma > 1\). (C) Dependence of the active bin density (\(\rho_{\text{bin}}\)) on the time bin size \(\Delta t\), for different values of \(\sigma\). Larger bin sizes lead to an increase in \(\rho_{\text{bin}}\), illustrating how binning affects activity measurements differently across different dynamical phases (D) Same analysis as in (C), but with $\Delta t$ chosen based on the mean interspike interval of the network, resulting in a nearly constant active bin density throughout the different phases.
}
    \label{fig:fig3}
\end{figure*}

\subsubsection{Surrogate data}
\label{sec:surrogate}

To test the robustness of our results, we compared them with surrogate data generated by shuffling the interspike intervals --- \textit{i.e.} the time series containing the time intervals between consecutive spikes --- of each unit within each time window (Fig.~\ref{fig:fig1}B). This process effectively breaks the correlations between units, producing trivial results with gaussian-distributed coarse-grained variables (Fig.~\ref{fig:fig1}D). 

\subsection{Adaptive time bin}\label{sec:methodsadaptive}

To account for the different activity levels across dynamical phases, we employ an adaptive time binning strategy. 
We choose larger time bins when activity is sparse and shorter time bins when activity is dense.

Specifically, we employ a time bin which is proportional to the average interspike interval, thus making our binning choice data-driven.
Given the average interspike interval $\langle \text{ISI} \rangle_j$ of neuron $j$, we calculate the population average

\begin{equation}
    \langle \text{ISI} \rangle = \frac{1}{N_{prg}} \sum_j^{N_{prg}} \langle \text{ISI} \rangle_j \; .
\end{equation}

Additionally, we can tune the time bin by means of a scaling factor $f$: 
\begin{equation}
\label{eq:tuning_facto}
    \Delta t = f \cdot \langle \text{ISI} \rangle \; ,
\end{equation}
so that $f$ is the only parameter governing data preprocessing. 
We will show that, as long as we choose $\Delta t$ according to Eq.~\eqref{eq:tuning_facto}, in such a way that $\rho_{\mathrm{bin}}$ becomes stable across dynamical phases the specific value of $f$ is not as important.

\section{Results}

\subsection{Timescale mismatch yields spurious signatures of criticality}\label{fixedbin}

First, we naively apply the PRG analysis to data from the excitable cellular automaton model using fixed time bins. 
This is a common heuristic choice in data analysis that does not consider the data structure or the distinct timescales across dynamical phases. As we will see, this can lead to spurious results. 

Consider, for instance, a relatively ``small'' time bin, $\Delta t = 25$~ms. 
If the model is in the supercritical regime (blue points in Fig.~\ref{fig:fig2}A), 
spikes are abundant enough so that there is enough information in the correlations among the $\varphi_i$'s. 
The resulting $\psi_i$'s are nearly gaussian-distributed, with low kurtoses, as expected from trivial (non-critical) behavior. 

In the subcritical regime, on the other hand, data is so sparse that the PRG incorrectly detects spurious correlations among the many zeros. 
The resulting coarse-grained activity distribution is sharply peaked around zero (Fig.~\ref{fig:fig2}B), which in turn leads to  high kurtoses. 
This can misleadingly suggest scale-invariance, despite the underlying dynamics being trivial.

Let us now turn our attention to a ``large'' time bin (say, ten times larger), $\Delta t = 120$~ms. 
Here, the situation is exactly reversed. 
Now the time bin is large enough to capture information in the sparse subcritical phase (black points in Fig.~\ref{fig:fig2}C), with Gaussian distributions and low kurtoses. 
In the supercritical phase, however, it is the silences that become too sparse. 
Once again, we have a sharply peaked distribution of coarse-grained activity (Fig.~\ref{fig:fig2} D) which, since the $\psi_i$'s are demeaned by construction, remains centered at zero [see Eq.~\eqref{eq:psi}]. 
We are once more left with a misleading signature of scale-invariance. 

This issue persists generically. 
As shown in Fig.~\ref{fig:widerfixedbin} (Appendix~\ref{app:fixedtimebin}), no single value of the time bin $\Delta t$ meaningfully accommodates both subcritical and supercritical phases of either model.

\subsection{Adaptive time bin reliably detects criticality}

\begin{figure*}[htpb]
    \centering
    \includegraphics[scale = 0.5]{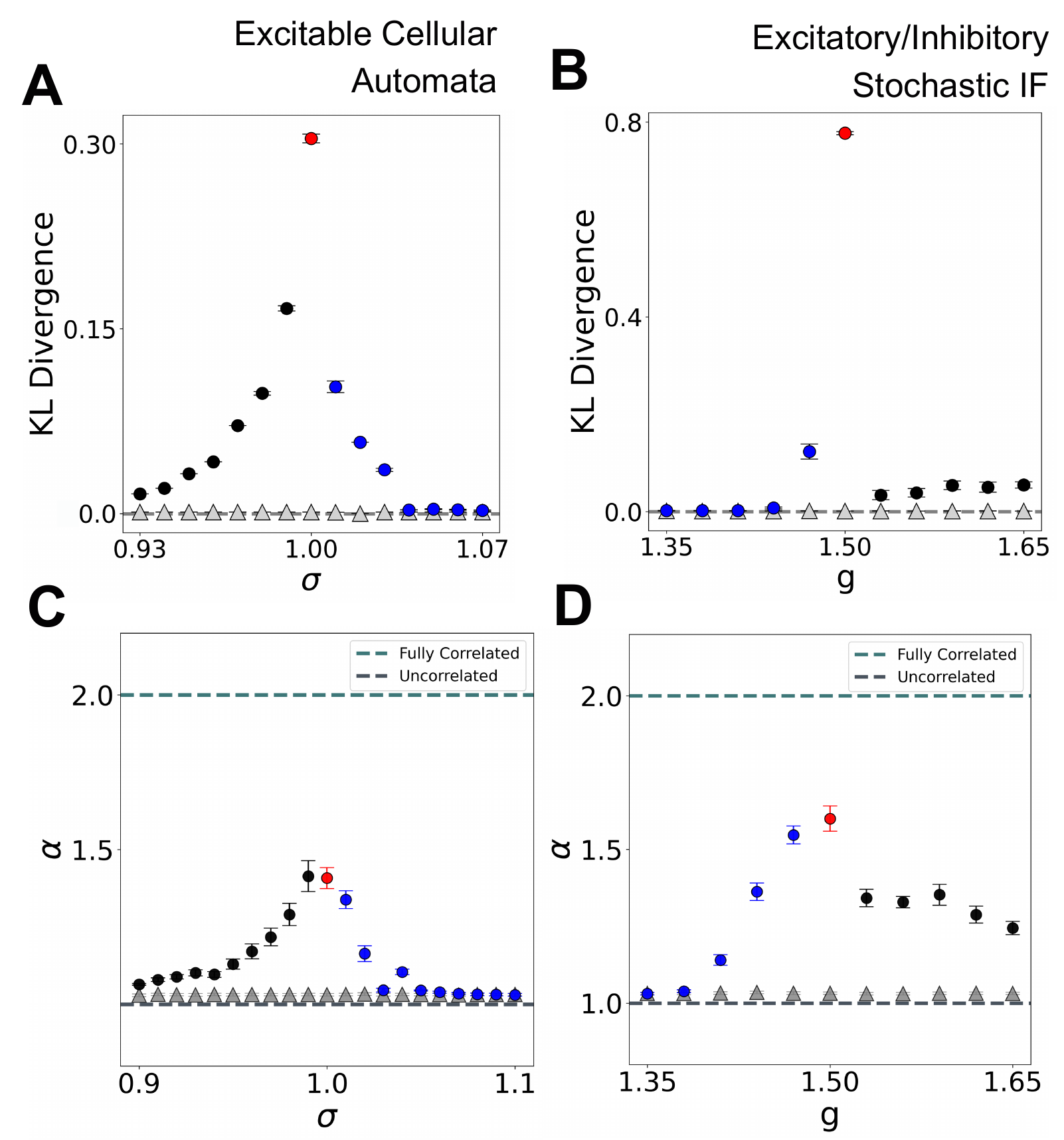} \caption{ Critical signatures of both neuronal models as their control parameters are swept across the critical point, with active bin density fixed ($f=0.26$ for the cellular automaton model, and $f=0.15$ for the stochastic integrate-and-fire model). (A,B) show results for the excitable cellular automaton model, while (C,D) correspond to the stochastic integrate-and-fire network. Circles show simulation data; triangles show shuffled controls. (A,C) Kullback-Leibler divergence (KL) peaks sharply at criticality (red), as the network goes from the subcritical (black) to the supercritical (blue) phase. (B,D) The exponent $\alpha$ also peaks around criticality, but more gradually. Shuffled data remains near the Gaussian regime throughout.
}
    \label{fig:fig4}
\end{figure*}

As we can see in Fig.~\ref{fig:fig3}A for three identical time series (therefore identical values of $\rho$), $\rho_{bin}$ increases with $\Delta t$. 
On its turn, $\rho$ depends on the model parameters (Fig.~\ref{fig:fig3}B). 
Intuitively, as a rule of thumb, one should avoid $\rho_{bin}$ too close to zero or too close to one. 
As we can see in Fig.~\ref{fig:fig3}C, which shows the joint effect of the collective state of the network and time binning on $\rho_{bin}$, no single fixed value of $\Delta t$ can solve this problem.

To address this, we introduce a data-driven approach to time binning that ensures a nearly constant $\rho_{bin}$ across different phases. 
As described in detail in Section~\ref{sec:methodsadaptive}, we adopt a solution which has appeared in the field of neuronal avalanches, namely, to choose a time bin proportional to the average interspike interval $\langle \text{ISI} \rangle$~\cite{Beggs03, Ribeiro10a, Capek2023parabolic, srinivasan2024recovery} as in eq.~\ref{eq:tuning_facto}. 
With this approach, we rely on the structure of the data itself to ensure that the chosen binning $\Delta t$ reflects an appropriate timescale for the dynamics under study. As a result, the active bin density remains nearly constant across different dynamical phases (Fig.~\ref{fig:fig3}D).
Therefore, the range of $f$ (from Eq.~\ref{eq:tuning_facto}) used throughout this work was chosen based on the requirement of maintaining an approximately constant active-bin density. 
A detailed justification of this choice, together with results for other values of $f$, is provided in appendix~\ref{app:isifactor}.

\begin{figure*}[htpb]
    \centering
    \includegraphics[scale = 0.5]{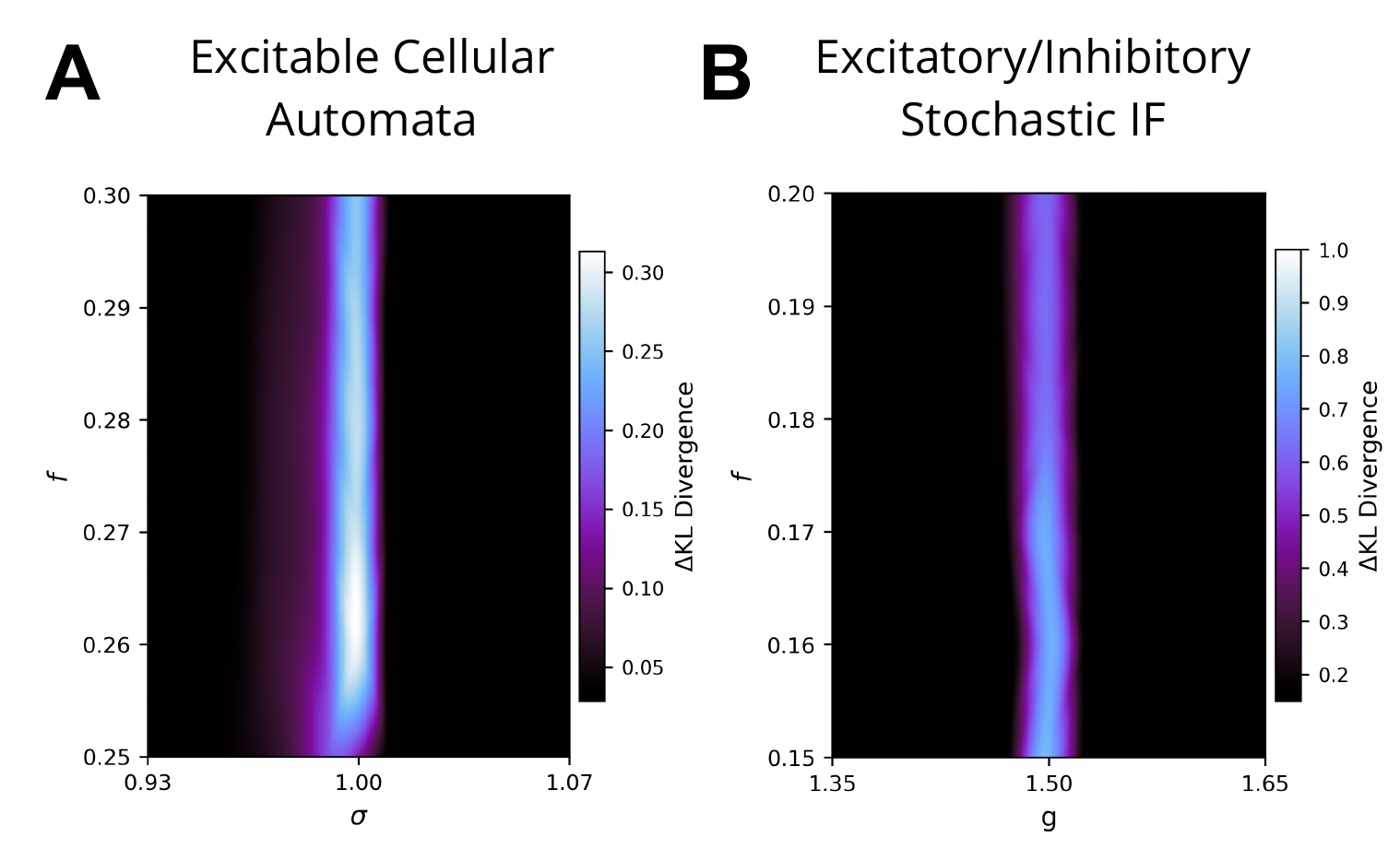} \caption{ 
    Difference between the  KL divergences of data and shuffled-data relative to a Gaussian distribution ($\Delta\mathrm{KL}$) across the control parameter and ISI-scaling factor $f$ for both (A) the excitable cellular automata model and (B) the stochastic E/I integrate-and-fire model. A narrow vertical ridge appears only in the vicinity of, and is centered around the critical point ($\sigma = 1$ for the excitable cellular automaton and $g = 1.5$ for the stochastic E/I IF model). The persistence of this fine ridge around the critical point across a broad band of $f$ shows that the scaling signature does not depend on fine–tuning the binning factor.}
    \label{fig:kurtosis heat}
\end{figure*}

Having established an appropriate binning strategy, we apply it to both the cellular automaton and the spiking model with excitation and inhibition.
In both cases, we vary the control parameters within ten percent of their critical values 
and calculate the Kullback–Leibler divergence (KL) of $P(\psi)$ with respect to a Gaussian, which provides a more information-theoretically grounded measure of non-Gaussianity than individual moments.
KL divergence increases significantly only near the critical point, highlighting its sensitivity to critical dynamics (Figs.~\ref{fig:fig4}A and~\ref{fig:fig4}C). 
We observe a sharp peak in the  KL divergence centered precisely at criticality. 
For a control parameter sufficiently distant from criticality, the KL divergence approaches zero, indicating trivial behavior. 
In contrast, the surrogate data remains consistently close to the trivial baseline across the entire parameter range. This behavior does not rely on selecting a finely tuned or “optimal’’ value of $f$; it persists across a range of tuning factors, as shown in Fig.~\ref{fig:kurtosis heat}. 
It is also observed for other metrics of deviations from a gaussian, such as skewness, kurtosis and fifth moment of the coarse-grained distributions (Fig.~\ref{fig:othermetrics}). 
This result reinforces the interpretation of deviation from Gaussianity of $P(\psi)$ as a signature of criticality when applying PRG to experimental data~\cite{cambrainha2025criticality, castro_and_2024}. 
In Appendix~\ref{app:urethane}, we illustrate how the adaptive binning procedure may be extended to the experimental recordings shown in Fig.~\ref{fig:fig1}. 
Although the parameter $f$ was introduced primarily to address methodological issues arising in the model analysis, its application to experimental data can also be considered for completeness.

\subsection{Mean variance exponent}

Next, we focus on the behavior of the $\alpha$ exponent obtained through real-space coarse graining (section~\ref{sec:realspace}).
As we sweep the control parameter, surrogate data remains close to the uncorrelated baseline ($\alpha = 1$), while actual data transitions into the non-trivial regime (${1 < \alpha < 2}$) near the critical point (Figs.~\ref{fig:fig4} B and~\ref{fig:fig4}D). 
In particular, the curves for $\alpha$ are broader and less precise than those of KL divergence. 
We also note that, for the model with excitation and inhibition (Fig.~\ref{fig:fig4}D), the nontrivial value of $\alpha$ extends into a larger interval of the subcritical regime. Notably, values found within this interval ($\alpha \sim 1.5$) were recently observed in empirical data for several different animal species, ranging from zebrafish and \textit{C. Elegans} to macaques~\cite{munn_multiscale_org}, which in the second model corresponds to an interval of the control parameter $g$ less than $4\%$ away from its critical value.

\section{Discussion}

The phenomenological renormalization group was proposed as a powerful tool for detecting scale invariance, its effectiveness coming precisely from its model independence~\cite{bradde_pca_2017, meshulam_coarse_2019}. 
A growing literature of applications of PRG to experimental data has since emerged, hinging on the assumption that the method’s signatures of scale invariance are a reliable marker of proximity to a critical point. 
Critical dynamics (as inferred with PRG) were claimed as an organizing principle not only across the whole brain~\cite{ponce2023critical, Morales2023scaling, castro2025interdependent}, but also across different species~\cite{munn_multiscale_org}. 
Recently, PRG was also employed to connect enhanced performance in a visual task 
with stronger signatures of criticality in the mouse visual cortex~\cite{cambrainha2025criticality}.

Our results provide the first systematic computational validation of this assumption in neuronal models across the critical region. 
We show that deviations from gaussianity in the distribution of normalized coarse-grained activity, measured through the kurtosis or other metrics, emerge only within a narrow vicinity of the critical point. 
These signatures are maximized at criticality, indicating that the PRG detects meaningful structure only in this regime.
We note that the models have been analyzed under heavy subsampling,  mirroring the limitations of experimental recordings. 
The models also allow us to check that these signatures are not a mere artifact of finite size (Appendix~\ref{app:finitesize}). 
Crucially, the model incorporating inhibition, a central feature of neuronal dynamics, exhibits even clearer critical signatures, further strengthening the relevance of comparing these results with PRG-based analyses of experimental data.

Furthermore, our work not only reinforces that the PRG is a powerful probe of critical dynamics, it also reveals that proper preprocessing is essential. 
In particular, the choice of temporal binning strongly affects the outcome: using an inappropriate timescale can introduce false signatures of scaling (Fig.~\ref{fig:fig2} and Fig.~\ref{fig:widerfixedbin}). 
By contrast, with our proposed binning scheme, based on the model’s intrinsic timescale, we observe scaling behavior only in the immediate vicinity of the critical point (Fig.~\ref{fig:kurtosis heat}), and this remains stable as long as the binning is adequate for surrogate data to yield trivial results. 
This provides a practical sanity check that isolates genuine critical signatures from artifacts introduced by inappropriate time binning.

Altogether, our findings establish a more solid foundation for the use of the PRG as a probe of critical dynamics in neuronal models and experimental data. In doing so, they reinforce the plausibility of earlier claims of criticality in the brain.

\begin{acknowledgments}

The authors acknowledge support from the Brazilian agencies CAPES (Grants No. 88887.900814/2023-00, No. 88887.021298/2024-00, No. PROEX 23038.003069/2022-87 and No. PROEX 1575/2024),
CNPq (Grants No. 140660/2022-4, No. 308703/2022-7 and No. 444500/2024-3), FADE/UFPE Grant No. 64/2024 and FACEPE
Grant No. IBPG-0934-1.05/25.
This research is supported
by INCT NeuroComp (CNPq Grant 408389/2024-9).

\end{acknowledgments}

\appendix

\section{Alternative measures of non-gaussianity}\label{alt_gauss}

\begin{figure*}[!t]
    \centering
    \includegraphics[width = 0.9\linewidth]{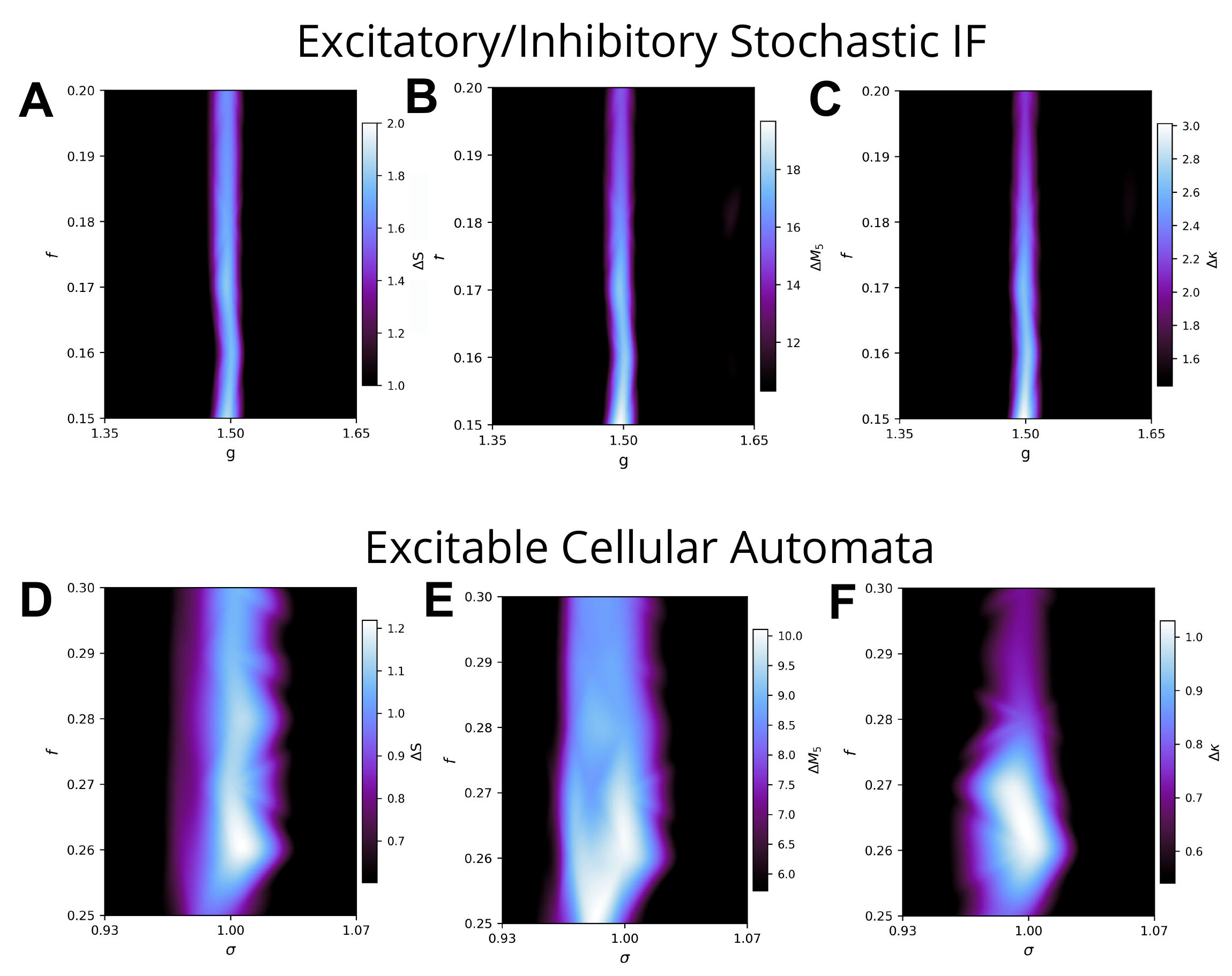}
    \caption{Alternative  measures of Gaussianity deviation for both models. (A–C) Stochastic integrate–fire network: heatmaps of the deviations in skewness $\Delta S$, the fifth standardized moment $\Delta M_{5}$ and kurtosis. The heatmaps all produce narrow ridges of non gaussianity localized around the critical coupling. (D–F) Excitable cellular automaton:  the same three metrics, all three exhibit broad but still well–defined ridges centered on the critical point. Together, these panels show that multiple independent measures of non-Gaussianity peak systematically at criticality in both models.}
    \label{fig:othermetrics}
\end{figure*}

For robustness, we now present additional proxies for deviations from Gaussian behavior, complementing the kurtosis-based analysis (Fig.~\ref{fig:othermetrics}). For a range of tuning parameter values, we compute the skewness 
${S = \langle \psi^3 \rangle / \langle \psi^2 \rangle^{3/2}}$ 
(Fig.~\ref{fig:othermetrics}A and~\ref{fig:othermetrics}D), the standardized fifth moment ${M_5 = \langle \psi^5 \rangle / \langle \psi^2 \rangle^{5/2}}$ (Fig.~\ref{fig:othermetrics}B and~\ref{fig:othermetrics}E), and kurtosis
(Fig.~\ref{fig:othermetrics}C and~\ref{fig:othermetrics}F) of the coarse-grained activity distribution. These quantities provide further diagnostics of asymmetry, higher-order fluctuations, and overall departures from gaussianity, which is our proxy for criticality. We observe that the trends shown in Fig.~\ref{fig:fig4} and Fig.~\ref{fig:kurtosis heat} for the KL divergence are also present for these other metrics.

\section{Fixed time bin}
\begin{figure}[!t]
    \centering
    \includegraphics[width = 0.95\columnwidth]{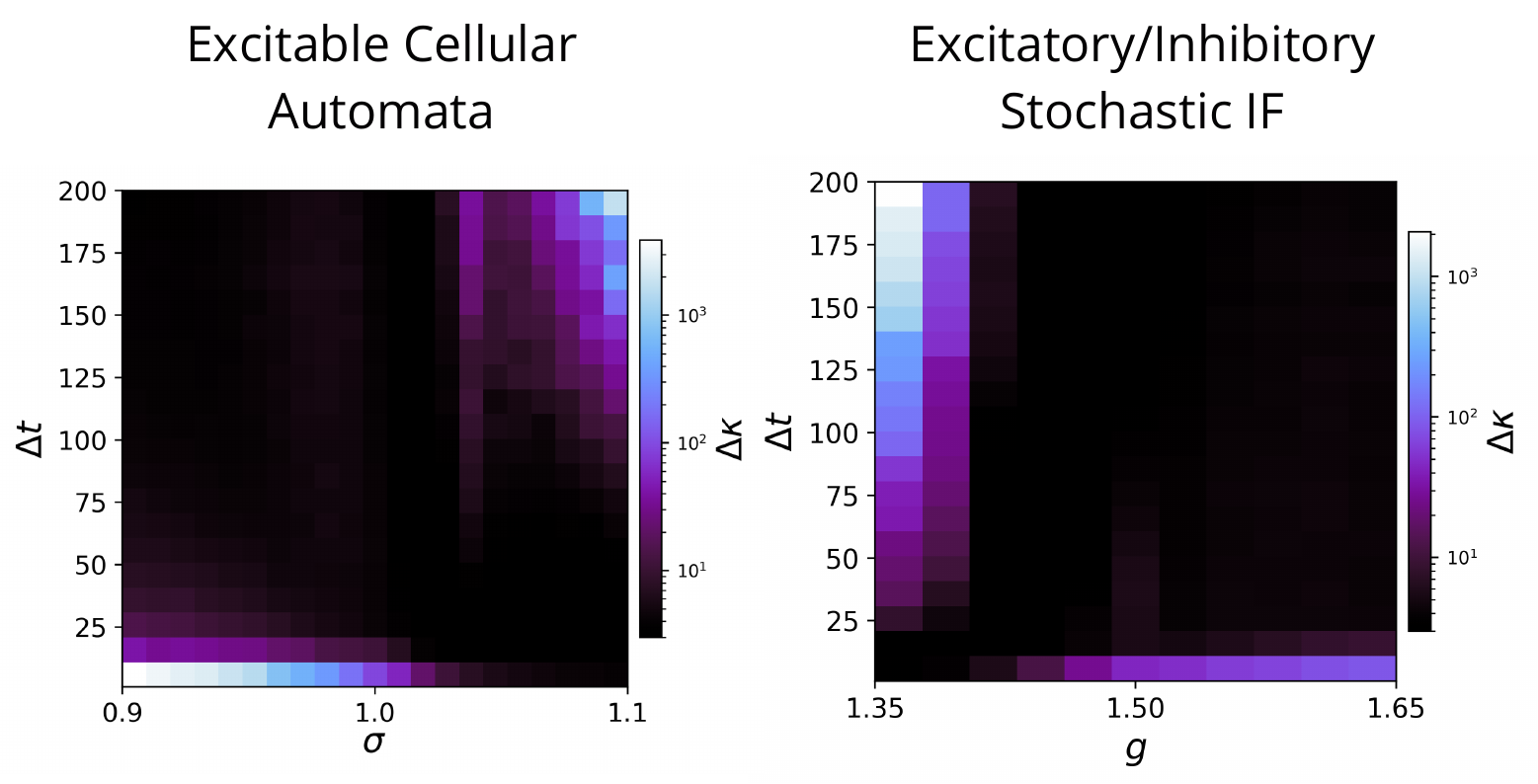} \caption{ Fixed bin analysis for both models. Heatmaps show $\Delta \kappa$ for a wider range of fixed bin sizes, across control parameters. (A) Shows results for the excitable cellular automata model, we notice areas of spurious high $\Delta \kappa$ for subcritical values (low $\sigma$)  and low bin values, and also for supercritical values (high $\sigma$) with large bin sizes. (B) Same trends as in (A), with small bins and subcritical values (high $g$) and large bins and supercritical values (low $g$) presenting spuriously high $\Delta \kappa$. 
}
    \label{fig:widerfixedbin}
\end{figure}

As discussed in Sec.~\ref{fixedbin}, spurious indicators of criticality emerge when the binning timescale is improperly chosen (Fig.~\ref{fig:fig2}). 
This effect persists across a wide range of bin widths and, as shown in Fig.~\ref{fig:widerfixedbin}, for sufficiently extreme values it produces clear false–scaling signatures in $\Delta\kappa$ (the kurtosis shift relative to shuffled data). 
Notably, no fixed timescale yields a $\Delta \kappa$ which peaks at the critical point and vanishes otherwise.

\label{app:fixedtimebin}

\section{Choice of $f$ parameter}\label{app:isifactor}
\begin{figure}[!t]
    \centering
    \includegraphics[width = 0.95\columnwidth]{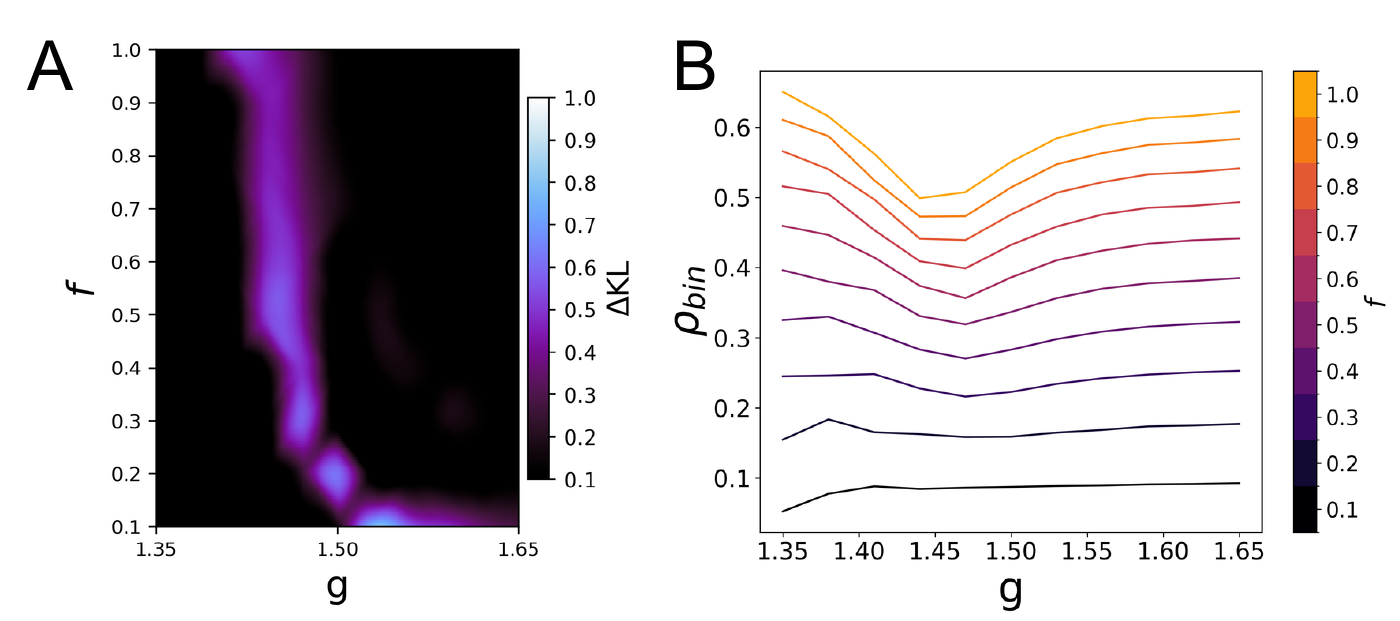} \caption{ (A) Difference between the KL divergences of the data and shuffled data relative to a Gaussian distribution ($\Delta \mathrm{KL}$) as a function of the control parameter $g$ and the ISI-scaling factor $f$ for the E/I stochastic IF model. The $\Delta \mathrm{KL}$ peak shifts slightly with increasing $f$, but remains within approximately $5\%$ of the critical value. 
(B) Average activity density ($\rho_{\mathrm{bin}}$) over the same parameter space. Note that for $f \sim 0.3$, the activity-density curve becomes unstable near the $g$ values where the shifted $\Delta \mathrm{KL}$ peak is observed in panel A.}
    \label{fig:f_range}
\end{figure}

When applying PRG to our models, some consistent method is required to compare data obtained in distant regions of parameter space and, therefore, exhibits vastly different levels of activity. 
When activity levels vary substantially, fixed bin sizes can introduce spurious correlations. 
The adaptive binning method mitigates this issue, allowing us to systematically analyze the emergence of PRG’s signatures of criticality as we sweep the control parameter.
The range of $f$ values was chosen to stabilize $\rho_\text{bin}$ across dynamical phases. 
Here we show that, for $f\lesssim 0.1$, the results for the E/I stochastic IF model in the subcritical regime become spurious because activity becomes too sparse (Fig.~\ref{fig:f_range}A). 
Also, in the supercritical regime the ISI becomes too small, and for $f \lesssim 0.1$ bins can become smaller than one time step of the model. 

At the other extreme, let us consider $f=1$.  
For a subcritical regime (e.g.  $g=1.65$), ISI values can become very large, leading to bin sizes that can exceed $\Delta t=900$~ms, which would be an unrealistic choice for real spiking data. 
In fact, already for $f\gtrsim 0.3$, $\rho_\text{bin}$ becomes progressively less stable (Fig.~\ref{fig:f_range}B), thereby defeating the very purpose of introducing the parameter. Despite these caveats, the peak in KL for larger values of $f$ is displaced from the true value by less than 5\% (Fig.~\ref{fig:f_range}A).

\section{Effects of adaptive bin size in data analysis}\label{app:urethane}

\begin{figure}[!ht]
    \centering
    \includegraphics[width=0.99\columnwidth]{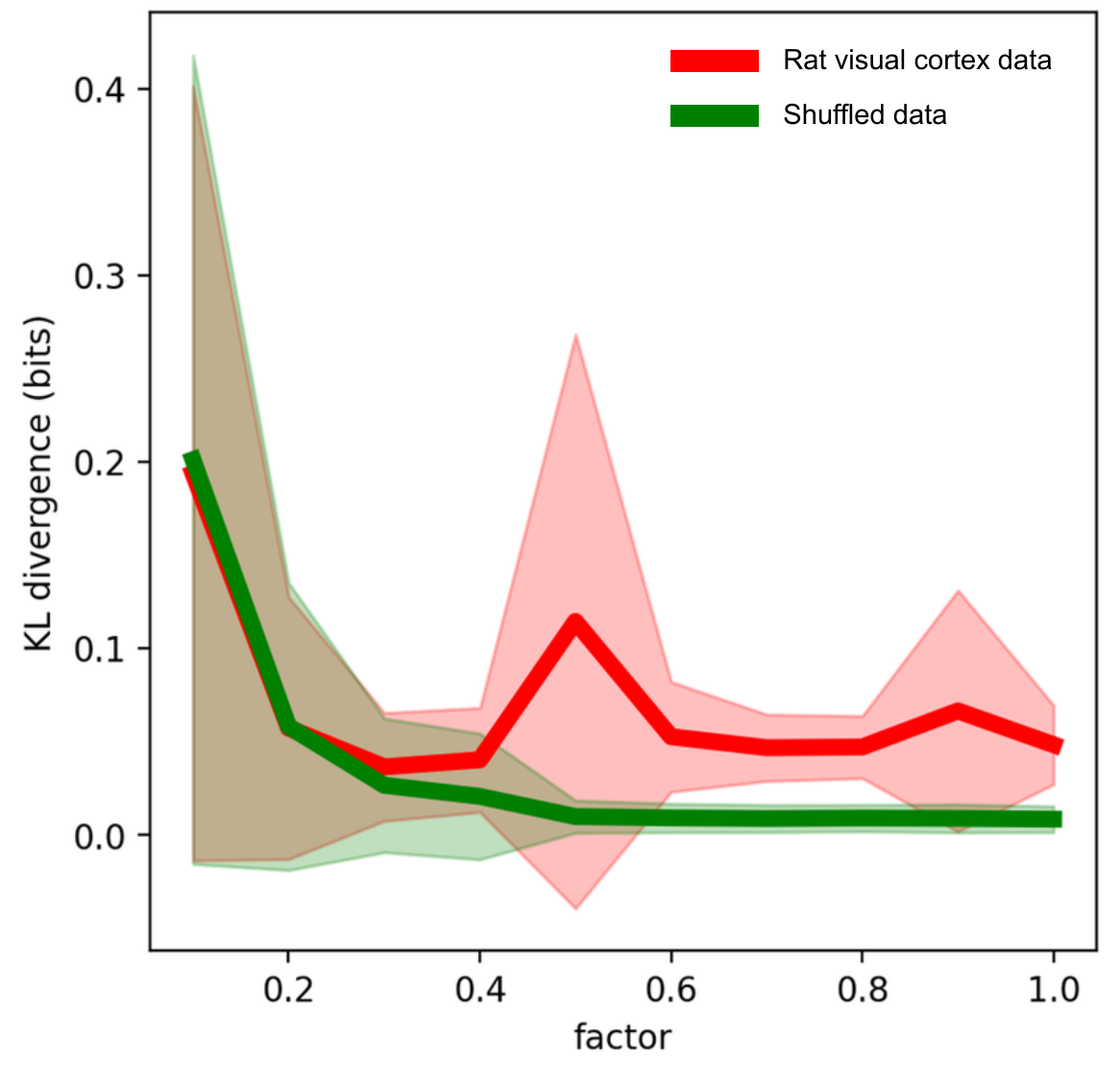} 
    \caption{KL divergence of the real (red) and shuffled (green) urethane-anesthetized rat V1 data with respect to a zero-mean unit-variance Gaussian, as a function of the scaling factor f, averaged across 9 rats. Solid lines represent the average and shaded areas represent the standard deviation.}
    \label{fig:urethaneffactor}
\end{figure}

We have applied the adaptive bin size to the spiking data from the primary visual cortex (V1) of urethane-anesthetized rats~\cite{castro_and_2024}.
For each of the 9 animals, data was timebinned based on the interspike interval as discussed in section~\ref{sec:methodsadaptive}, partitioned into time windows of 600 bins, then coarse-grained following the methods discussed (section~\ref{sec:momentumrenorm}). Employing the Kullback-Leibler divergence relative to a gaussian distribution as a measure of non-gaussianity, two complementary results can be observed in Fig.~\ref{fig:urethaneffactor}. 
First, that for vanishing values of $f$ and therefore very small bin sizes, we observe spurious correlations due to excessive silence in both the experimental data and its shuffled counterpart. 
In other words, even without a priori knowledge of ground-truth criticality from a model, the data itself can reveal that $f$ is too small by yielding non-gaussian $P(\psi)$ even for shuffled data. 
Second, that as soon as $f$ is large enough, the KL divergence for real data and shuffled data separate, with shuffled-data staying near zero, while real data remains non-gaussian. 
This is in agreement with previous results suggesting that primary visual cortex spiking data of urethane-anesthetized rats exhibits signatures of criticality~\cite{Fontenele2019criticality,Carvalho2021subsampled,lotfi2020signatures,lotfi2021statistical,castro_and_2024}.

\section{Effects of system size}
\label{app:finitesize}

\begin{figure}[!t]
    \centering
    \includegraphics[scale = 0.68]{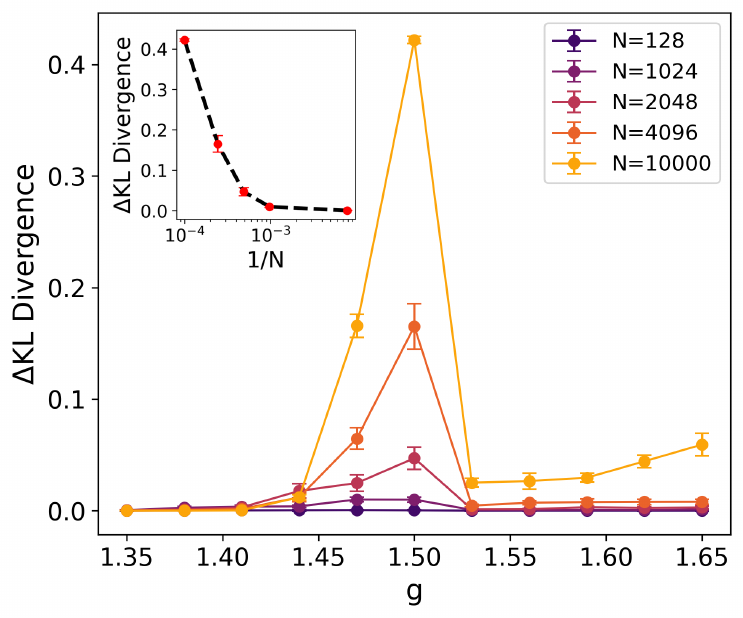} 
    \caption{ Dependence of the difference of KL divergence of data and shuffled data, on the control parameter $g$ for the stochastic excitatory/inhibitory integrate--and--fire model at different system sizes $N$. The peak at the critical value $g \approx 1.5$ grows systematically with $N$, indicating that deviations from Gaussianity strengthen with system size. Inset: $\Delta KL$ at the critical point plotted against $1/N$, showing stronger deviations from gaussianity in the probability distribution of coarse-grained normalized activity as system size is increase.}

    \label{fig:systemsizeanalaysis}
\end{figure}

We test how the KL divergence of the distribution of coarse-grained activity depends on system size (Fig.~\ref{fig:systemsizeanalaysis}). The quantity analyzed here is the kurtosis of the final distribution of coarse-grained activity, which is distinct from the order-parameter distribution commonly considered in equilibrium models.

For each system size, the KL divergence is computed across a range of tuning parameters. The results show a systematic trend: the peak becomes sharper and its maximum increases as the system size grows (Fig.~\ref{fig:systemsizeanalaysis}). Larger systems display a more pronounced maximum (Fig.~\ref{fig:systemsizeanalaysis}, inset). This indicates that the signatures of criticality observed with PRG analysis intensify as we increase  system size and are not an artifact of small systems.

\bibliography{copelli1}
\bibliographystyle{apsrev4-2}

\end{document}